\documentclass{article}
\usepackage{epsfig,amsmath,amssymb,enumitem}
\usepackage[papersize={8.5in,11in}]{geometry}
\usepackage{multirow}
\begin{document}
\title{Quarkonium resonance identified with the 125 GeV boson}
\author{J. W. Moffat\\~\\
Perimeter Institute for Theoretical Physics, Waterloo, Ontario N2L 2Y5, Canada\\
and\\
Department of Physics and Astronomy, University of Waterloo, Waterloo,\\
Ontario N2L 3G1, Canada}
\maketitle
\begin{abstract}
The 125 GeV resonance discovered at the LHC could be a heavy quarkonium, spin 0 pseudoscalar meson $\zeta^0$. The decay rates of the $\zeta^0$ meson resonance are calculated and compared to the standard model Higgs boson decay rates. The branching ratios and signal strengths for  $\zeta\rightarrow\gamma\gamma$, $\zeta\rightarrow Z\gamma$, $\zeta\rightarrow ZZ^*$ and $\zeta\rightarrow WW^*$ are approximately the same as the Higgs boson branching ratios and signal strengths. The decay rates for $\zeta\rightarrow\tau^+\tau^-$, $\zeta\rightarrow b\bar b$ and $\zeta\rightarrow c\bar c$ are suppressed compared to the Higgs boson decay rates. Accurate branching ratios and signal strengths obtained at the LHC can distinguish between the standard model Higgs boson and the heavy composite $\zeta$ meson resonance.
\end{abstract}

\section{Introduction}

The Large Hadron Collider (LHC) has discovered a new boson. The CMS~\cite{CMS} and ATLAS~\cite{ATLAS} experiments based on $5\,{\rm fb}^{-1}$ data show an excess of events at $\sim 125-126$ GeV that was evident already in the 2011 data. We must make certain that the new boson is the standard model Higgs boson that was postulated to generate electroweak symmetry breaking, which gives masses to the standard model vector bosons $W$ and $Z$ and to the fermions~\cite{Djouadi}. The LHC experiments will eventually have the sensitivity to accurately determine the decay rates for the final states $\gamma\gamma, ZZ^*, WW^*,Z\gamma,b\bar b, c\bar c, gg$ and $\tau^+\tau^-$ predicted by the Higgs boson production mechanism. The excess of events at $\sim 125$ GeV is mainly driven by the ``golden'' decay channels $\gamma\gamma$ and $ZZ^*\rightarrow 4$ leptons. The decay channels $H\rightarrow ZZ^*$ and $H\rightarrow WW^*$ are significant signatures of the Higgs coupling to the vector bosons. Because the standard model elementary Higgs boson is necessarily a scalar boson, $J^{PC}=0^{++}$, with even parity, {\it it is critically important to determine the parity of the new boson}. In particular, if the boson with spin 0 has negative parity i.e., it is a pseudoscalar boson, then this has significant consequences for the standard model; the theory is not gauge invariant rnormalizable in its couplings to the $W$ and $Z$ bosons. A pseudoscalar Higgs boson will predict a decay rate $\Gamma(H\rightarrow ZZ^*\rightarrow 4\ell)$ that is significantly suppressed compared to the standard model Higgs boson prediction, and will not agree with the $3\sigma$ result obtained by the CMS and ATLAS groups. The spin and parity of the new boson can be determined at the LHC by analyzing the correlations of the angular distributions of the $\gamma\gamma$ and 4-lepton decay states of the 125 GeV boson~\cite{Choi,Gao,Rujula,Rujula2,Logan,Stolarski,Ellis,Gao2,Drozdetskiy}.

Recently, we proposed a model of a pseudoscalar meson resonance called $\zeta^0$, which is a composite mixture of bottomonium and toponium~\cite{Moffat,Moffat2}.  A detailed derivation of the partial decay widths of the $\zeta$ meson is compared to the predicted decay widths of the standard model Higgs boson. The branching ratios and signal strengths for the $\zeta$ meson decays into bosons are comparable in size to the Higgs boson decays, but the fermion-antifermion decay rates are suppressed compared to the Higgs boson. The Moriond 2013 conference has updated the data for the decay channels of the new boson $X$ and found that the standard Higgs boson model is compatible with the data~\cite{Hubaut,Mountricha,VMartin,Dutta,Puigh}. However, the fermion-antifermion decay channels $X\rightarrow \tau^+\tau^-, b\bar b$ and $c\bar c$ reported by the ATLAS collaboration correspond to a null signal within $1\sigma$~\cite{VMartin,Puigh}, while the CMS results~\cite{Dutta,Puigh} are marginally consistent with the standard Higgs boson model predictions.

\section{The 125 GeV resonance as a mixture of bottomonium and toponium}

We cannot picture the $\zeta$ resonance as a quarkonium meson composed of a quark-anti-quark bound state formed from a fourth generation of quarks $(t',b')$ with $m_{b'}\sim 62.5$ GeV and $m_\zeta\sim 2m_{b'}\sim 125$ GeV, because such a fourth generation of quarks has not been found. A relatively light $b'$ quark would be copiously produced in associated production, $p+p^{\pm}\rightarrow b'+{\rm anything}$ as is the case with the b quark. Moreover, the semi-leptonic decay of a $b'$ quark with a life-time comparable to the $b$ quark would have been detected at the Tevatron and the LHC~\cite{pdg}. Instead, we assume that the $\zeta$ is a bound state mixture of bottomonium and toponium.

The bottomonium and unobserved toponium are isoscalar states $\vert B\rangle=\vert b{\bar b}\rangle$ and $\vert T\rangle=\vert t{\bar t}\rangle$ of heavy quarkonium. With respect to an effective interaction Hamiltonian, heavy quarkonium resonances appear in two different isoscalar states $\vert\zeta^0\rangle$ and $\vert\zeta^{0'}\rangle$. The effective Hamiltonian is given by~\cite{Moffat,Moffat2}:
\begin{equation}
{\cal H}_{\rm eff}={\cal H}_0+{\cal H}_{\rm mass},
\end{equation}
where
\begin{equation}
{\cal H}_{\rm mass}=K^T{\cal M}K.
\end{equation}
Here, ${\cal M}$ is the mass matrix:
\begin{equation}
\label{Massmatrix}
{\cal M}=\biggl(\begin{array}{cc}m_{\zeta'}&m_{\zeta\zeta'}\\
m_{\zeta\zeta'}&m_{\zeta}\end{array}\biggr),
\end{equation}
where $\vert\zeta\rangle$ and $\vert\zeta'\rangle$ are states of quarkonium that interact through the mixing contributions $m_{\zeta\zeta'}$ and $K=\biggl(\begin{array}{cc}\zeta'\\\zeta\\\end{array}\biggr)$.
The mass matrix can be diagonalized:
\begin{equation}
\label{Diagonalmatrix}
D=R^T{\cal M}R,
\end{equation}
where $R$ is the rotation matrix:
\begin{equation}
\label{rotationmatrix}
R=\biggl(\begin{array}{cc}\cos\phi&\sin\phi\\-\sin\phi&\cos\phi\end{array}\biggr),
\end{equation}
and $D$ is the diagonal matrix:
\begin{equation}
D=\biggl(\begin{array}{cc}m_T&0\\0&m_B\end{array}\biggr).
\end{equation}
Here, $m_T\sim 2m_t\sim 346$ GeV and $m_B\sim 2m_b\sim 9$ GeV where we have used the measured quark masses: $m_t\sim 173$ GeV and $m_b\sim 4.5$ GeV.

In terms of the states $\vert T\rangle$ and $\vert B\rangle$, we have
\begin{equation}
{\cal H}'_{\rm mass}=m_TTT^\dagger+m_BBB^\dagger,
\end{equation}
where
\begin{equation}
\biggl(\begin{array}{cc}T\\B\\\end{array}\biggr)
=R\biggl(\begin{array}{cc}\zeta'\\\zeta\\\end{array}\biggr)
=\biggl(\begin{array}{cc}\cos\phi \zeta'+\sin\phi \zeta\\\cos\phi\zeta-\sin\phi\zeta'\end{array}\biggr).
\end{equation}
From (\ref{Diagonalmatrix}), we obtain the masses:
\begin{equation}
m_T=\cos^2\phi m_{\zeta'}-2\sin\phi\cos\phi m_{\zeta\zeta'}+\sin^2\phi m_\zeta,
\end{equation}
and
\begin{equation}
m_B=\cos^2\phi m_\zeta+2\sin\phi\cos\phi m_{\zeta\zeta'}+\sin^2\phi m_{\zeta'}.
\end{equation}

By inverting (\ref{Diagonalmatrix}), we get
\begin{equation}
{\cal M}=RDR^T
\end{equation}
which leads to the result
\begin{equation}
\label{zetasolution}
{\cal M}\equiv\biggl(\begin{array}{cc}m_{\zeta'}&m_{\zeta\zeta'}\\
m_{\zeta\zeta'}&m_\zeta\end{array}\biggr)=\biggl(\begin{array}{cc}\cos^2\phi m_T+\sin^2\phi m_B&\cos\phi\sin\phi(m_B-m_T)\\
\cos\phi\sin\phi(m_B-m_T)&\cos^2\phi m_B+\sin^2\phi m_T\end{array}\biggr).
\end{equation}
We derive from (\ref{zetasolution}) the results
\begin{equation}
\label{zetamass}
m_\zeta=\cos^2\phi m_B+\sin^2\phi m_T,
\end{equation}
and
\begin{equation}
\label{zetaprimemass}
m_{\zeta'}=\cos^2\phi m_T+\sin^2\phi m_B.
\end{equation}
The off-diagonal term is given by
\begin{equation}
m_{\zeta\zeta'}=\cos\phi\sin\phi(m_B-m_T).
\end{equation}

We can determine the mixing angle $\phi$ from the equation:
\begin{equation}
\label{mixingangle}
\phi=\arccos[(m_T-m_{\zeta})/(m_T-m_B)]^{1/2}.
\end{equation}
The mixing angle $\phi\sim 36\,^{\circ}$ is obtained from (\ref{mixingangle}) and from (\ref{zetamass}) and (\ref{zetaprimemass}), we get the masses of the quarkonium states $\vert\zeta\rangle$ and $\vert\zeta'\rangle$: $m_{\zeta^0}\sim 125$ GeV and $m_{\zeta^{0'}}\sim 230$ GeV. We identify the new boson resonance discovered at the LHC with the $\zeta^0$ bound state quarkonium resonance. The mixing of the $\zeta$ and the $\zeta'$ mesons is strongly enhanced by a non-perturbative instanton gluon interaction mediated by an axial current $U_A(1)$ anomaly~\cite{Moffatanomaly}.

We have used the zero width approximation to determine the diagonalization of the mass matrix and the masses of the $\zeta$ and $\zeta'$ resonances. We could have determined the mixing of the mass matrix using the finite resonance width or overlapping resonance formalism~\cite{Dothan,Dothan2,Gilman}. The $2\times 2$ mass matrix then takes the form:
\begin{equation}
\label{BRmassmatrix}
{\cal M}_0=\biggl(\begin{array}{cc}m_{0\zeta'}-i\Gamma_{0\zeta'}&\delta m\\
\delta m&m_{0\zeta}-i\Gamma_{0\zeta}\end{array}\biggr).
\end{equation}
Here, ${\cal M}_0$ is the undiagonalized mass matrix with elements expressed in terms of bare masses and widths $m_0$ and $\Gamma_0$ and the off-diagonal term $\delta m$, which induces the mixing, originates in the interaction of the $\zeta$ and $\zeta'$ mesons.

\section{Quarkonium potential models}

The decay widths and production cross sections of quarkonium states can be calculated from quarkonium wave functions~\cite{Appelquist,Novikov,Eichten,Barger,Kwong,Martin,REllis}. The calculation of the wave functions of superheavy quarkonium states requires an extrapolation from the observed charmonium and bottomonium regimes, so the results can be sensitive to the interquark potential model. By integrating the non-relativistic Schr\"odinger wave equation by parts and applying the normal boundary conditions at $r=0$ and $r=\infty$, we can relate the quarkonium wave functions squared at the origin to the derivatives of the potential:
\begin{equation}
\vert R_S(0)\vert^2=m_q\langle dV/dr\rangle,
\quad \vert R'_P(0)\vert^2=\frac{m_q}{9}\langle(dV)/(r^2dr)+4(E-V)/r^3)\rangle,
\end{equation}
where $\langle...\rangle$ denotes an expectation value, $E=m_{Q}-2m_q$, $Q$ denotes the quarkonium state and $R_s(0)$ and $R'_P(0)$ denote the radial wave function of the S-state and the derivative of the radial wave function of the P-wave state, respectively, evaluated at $r=0$. The quarkonium wave functions at the origin are mainly determined by the size of the quarkonium state, which is approximately the size of the Bohr radius $(\alpha_sm_q)^{-1}\sim 8/m_q$ where $\alpha_s$ is the strong coupling constant.

For the heavy quarkonium the short-distance part of the potential is dominated by the Coulomb potential:
\begin{equation}
\label{Coulomb}
V(r)=-\frac{4}{3}\frac{\alpha_s(m_q^2)}{r}.
\end{equation}
The Richardson potential~\cite{Richardson}:
\begin{equation}
V(q^2)=-\frac{4}{3}\frac{12\pi}{33-2N_f}\frac{1}{q^2}\frac{1}{\ln(1+q^2/\Lambda^2)}.
\end{equation}
incorporates the asymptotically free short distance behavior and a linear confinement potential in momentum space.
This is to be compared with the Cornell potential~\cite{Cornell}:
\begin{equation}
V(r)=-\frac{k}{r}+ar.
\end{equation}
The wave functions can be calculated for the lowest $S$ and $P$ states from the potential (\ref{Coulomb}):
\begin{equation}
\vert R_S(0)\vert^2=4\biggl(\frac{2}{3}\alpha_sm_q\biggr)^3=\frac{4}{27}\alpha_s^3m_Q^3,\quad \vert R'_P(0)\vert^2=\frac{32}{5832}\alpha_s^5m_Q^5,
\end{equation}
where $m_Q$ denotes the quarkonium mass.

If the life-time of the quarkonium resonance state is shorter than
\begin{equation}
\label{zetalifetime}
t_R\sim \frac{9}{2m_Q\alpha_s^2}\sim 1.7\times 10^{-24}\,s,
\end{equation}
then the quarkonium resonance cannot form a bound state~\cite{Bigi}. Here, $\alpha_s(m_Z)=0.118$ is the QCD coupling constant~\cite{pdg}.  This is the case for toponium due to the rapid decay of the top quark, $t\rightarrow bW^+$. We shall find that the decay life-times of $\zeta$ for $m_Q=m_\zeta\sim 125$ GeV are longer than (\ref{zetalifetime}), so that the $\zeta$ resonance can form a quasi-stable bound state. The heavy quark bound state zero point momentum is, $p\sim\alpha_sm_q$, yielding a binding energy of order
$E_{\rm bind}=p^2/2m_q\sim\alpha_s^2m_q$.

\section{Effective Lagrangians for $^1S_0$ decay into two photons}

An effective Lagrangian for the coupling of $q\bar q$ to two photons is given by~\cite{Lansberg}:
\begin{equation}
{\cal L}_{\gamma\gamma{\rm eff}}
=-ic_1(\bar{q}\gamma_\sigma\gamma^5q)\epsilon^{\mu\nu\rho\sigma}F_{\mu\nu}A_\rho.
\end{equation}
Here, the coupling constant $c_1$ is
\begin{equation}
c_1\sim \frac{e_q^2(4\pi\alpha)}{m_Q^2+b_Qm_Q},
\end{equation}
where $e_q$ is the quark charge per unit proton charge and $b_Q$ is the binding energy of the quarkonium state. The factor $1/(m_Q^2+b_Qm_Q)$ arises from the quark propagator:
\begin{equation}
\Delta_q=\frac{1}{(k_1-k_2)^2/4-m_q^2},
\end{equation}
where $k_1,k_2$ are the outgoing photon momenta.

We define
\begin{equation}
\langle 0\vert\bar q\gamma^\mu\gamma^5\vert Q\rangle=if_Qp^\mu,
\end{equation}
where $f_Q$ is the quarkonium decay constant. We obtain the following expression for the amplitude for $Q\rightarrow \gamma\gamma$:
\begin{equation}
{\cal M}_{\gamma\gamma}=-4ie_q^2(4\pi\alpha)\frac{f_Q}{m_Q^2+b_Qm_Q}\epsilon^{\mu\nu\rho\sigma}
\epsilon_{1\mu}\epsilon_{2\nu}k_{1\rho}k_{2\sigma}.
\end{equation}
This yields the partial quarkonium $Q(^1S)$ decay rate with $b_Q\sim 0$:
\begin{equation}
\Gamma(Q\rightarrow\gamma\gamma)=\frac{1}{2}\frac{1}{64\pi^2m_Q}\int d\Omega\vert{\cal M}\vert^2=\frac{4\pi e_q^4\alpha^2f_Q^2}{m_Q},
\end{equation}
where the factor $1/2$ is the Bose symmetry factor.

If we use $f_Q^2=12\vert R_S(0)\vert^2/m_Q$, we recover the non-relativistic result of Novikov et al., Barger et al., and Kwong et al.~\cite{Novikov,Barger,Kwong}.\begin{footnote}{The formulas for the partial widths we use differ from the results of Barger et al.,~\cite{Barger} by a factor $4\pi$ in agreement with the results obtained by Kwong et al.,~\cite{Kwong} and Lansberg and Pham~\cite{Lansberg}.}\end{footnote}

\section{Decay and production rates of $\zeta$ resonance and Higgs boson}

The new boson at 125 GeV is observed to decay into two photons, so by the Landau-Yang theorem it cannot have spin $J=1$~\cite{Landau,Yang}. For a quarkonium spectrum, we can have the states $J^{PC}=0^{-+}, J^{PC}=1^{--}, J^{PC}=0^{++}, J^{PC}=1^{++}, J^{PC}=2^{++}$ and $J^{PC}=1^{+-}$, and only the spins $J=0,2$ can decay into the $\gamma\gamma$ and $gg$ channels (the Landau-Yang theorem is applicable to the two-gluon state, for the initial state is a color singlet.) We will assume that only the pseudoscalar $\zeta(0^{-+})$ quarkonium state at 125 GeV is presently observable.

The decays of the quarkonium state $\zeta\rightarrow p_1p_2$, where $p_1,p_2$ are a fermion-antifermion pair or a gauge boson pair, are described by diagrams corresponding to decays via the exchange of $\gamma, Z^0$ in the s channel, and decays via a quark exchange in the t or u channel.

We will compare the decay rates of the $\zeta$ resonance with the decay rates of the standard model Higgs boson with $J^{PC}=0^{++}$. The decay rates and branching ratios of the Higgs boson are obtained from the CERN working group and displayed in Table 1~\cite{Ellis2,Bergstrom,Kane,Steinhauser,CERNworkinggroup}.
$$ $$
1. Decay channels $\gamma\gamma, gg, Z\gamma$
$$ $$
The decay width for the channel $\zeta\rightarrow\gamma\gamma$ is given by
\begin{equation}
\label{zetawidth}
\Gamma(\zeta\rightarrow\gamma\gamma)=\frac{48\pi\alpha^2e_q^4}{m_\zeta^2}\vert R_S(0)\vert^2=\frac{192\pi}{27}\alpha^2\alpha_s^3e_q^4m_\zeta,
\end{equation}
where $\alpha=e^2/4\pi$.  Because the top quark charge dominates our mixture of $b$ and $t$ quarks that describe the $\zeta$ resonance, we choose $e_q=2/3$ in units of the proton charge. We obtain the decay rate
\begin{equation}
\Gamma(\zeta\rightarrow\gamma\gamma)=48.3\,{\rm keV}.
\end{equation}

The partial width of the $\zeta$ resonance decay into two gluons is given by
\begin{equation}
\Gamma(\zeta\rightarrow gg)=\frac{8\pi\alpha_s^2}{3m_\zeta^2}\vert R_S(0)\vert^2
=\frac{128\pi}{81}\alpha_s^5m_\zeta=14.2\,{\rm MeV}.
\end{equation}

For the case of the radial excitations of $\zeta\rightarrow\gamma\gamma$, we have~\cite{Martin}:
\begin{equation}
\vert R_P'(0)\vert^2=\frac{4}{n^3}\biggl(\frac{1}{3}\alpha_s^3m_\zeta\biggr),
\end{equation}
where $n$ is the quarkonium radial number. For $n=2$ we get
\begin{equation}
\Gamma(\zeta\rightarrow\gamma\gamma)=6\,{\rm keV}.
\end{equation}

The diagrams that occur for the decay $\zeta\rightarrow Z\gamma$ are the same as those for the two photon decay. The decay rate is given by~\cite{Barger}:
\begin{equation}
\Gamma(\zeta\rightarrow Z\gamma)=\frac{96\pi\alpha\alpha_Ze_q^2V_Q^2(1-R_Z)\vert R_S(0)\vert^2}{m_\zeta^2}=\frac{1536\pi\alpha\alpha_Z\alpha_s^3V_Q^2(1-R_Z)m_\zeta}{243},
\end{equation}
where $V_Q$ is the vector coupling of $Q$ to the $Z^0$:
\begin{equation}
V_Q=\frac{1}{3}(I_{3L}+I_{3R})-e_q\sin^2\theta_w.
\end{equation}
For fermions $I_{3L}(I_{3R})$ is the third component of the weak isospin for the left-(right-) handed fermion. We have $I_{3L}=\pm 1/2$ and $I_{3R}=0$ and we use $I_3=-1/2$ giving $V_Q=0.402$. Moreover, $\alpha_Z=\alpha/\sin^2\theta_w\cos^2\theta_w$ with $\sin^2\theta_w=0.231$ and $R_Z=m_Z^2/m_\zeta^2$. We obtain
\begin{equation}
\Gamma(\zeta\rightarrow Z\gamma)=92.7\,{\rm keV}.
\end{equation}

2. Decay channels $ZZ^*, WW^*$
$$ $$
The diagrams contributing to the $ZZ^*$ decay are quark exchanges in the t- and u-channels, as in the two photon decay. The longitudinal $Z_LZ_L^*$ decay is absent for $J^{PC}=0^{-+}$. The decay width for $\zeta\rightarrow ZZ^*$ when one or both of the $Z$s are off the mass-shell is given by~\cite{Barger}:
\begin{equation}
\Gamma(\zeta\rightarrow ZZ^*)=48\pi\alpha_Z^2(V_Q^2+A_Q^2)\tilde\beta_Z^3\frac{\vert R_S(0)\vert^2}
{m_\zeta^2(1-2{\tilde R}_Z)^2}=\frac{192\pi\alpha_Z^2\alpha^3_s(V_Q^2+A_Q^2)\tilde{\beta}_Z^3m_\zeta}{27(1-2\tilde{R}_Z)^2}.
\end{equation}
Here, $\tilde{R}_Z=m_1m_2/m_\zeta^2 < m_Z^2/m_\zeta^2$, where $m_1=m_Z\sim 91.18\,{\rm GeV}, m_2= m_{Z^*}\sim 33.8\,{\rm GeV}$, and $\tilde\beta_Z=(1-4\tilde{R}_Z)^{1/2}$. Moreover, the axial vector coupling to $Z^0$ is $A_Q=(I_{3L}-I_{3R})/2$. As before, we have $I_{3L}=\pm 1/2$, $I_{3R}=0$, $A_Q\sim 0$ and this yields $V_Q^2+A_Q^2\sim 0.162$. We obtain for the partial width:
\begin{equation}
\Gamma(\zeta\rightarrow ZZ^*)=331\,{\rm keV}.
\end{equation}

Only one t-channel quark exchange diagram occurs for the $W^+W^-$ decay. The s-channel exchanges of $\gamma$ and $Z$ are absent for $J^{PC}=0^{-+}$. The one quark exchange in the t-channel depends on the mass of the quark. The double longitudinal state $W^+_LW^-_L$ decay is absent for $J^{PC}=0^{-+}$. For the decay $\zeta\rightarrow W^+W^-$, when one or both of the $W$s are off the mass-shell, we obtain the partial decay width:
\begin{equation}
\Gamma(\zeta\rightarrow WW^*)=\frac{12\pi\alpha^2_W\tilde\beta_W^3\vert R_S(0)\vert^2}{8m_\zeta^2(1-Y)^2}
=\frac{48\pi\alpha^2_W\alpha_s^3\tilde\beta_W^3m_\zeta}{216(1-Y)^2}.
\end{equation}
Note the lack of Bose symmetry for $W^+W^-$ compared to $Z^0Z^0$.   We have $\alpha_W=\alpha/\sin^2\theta_w, \tilde\beta_W=(1-4\tilde{R}_W)^{1/2}, \tilde{R}_W=m_1m_2/m_\zeta^2$, and $m_1\sim  m_W= 80.39\,{\rm GeV}, m_2\sim m_{W^*}\sim 44.9\,{\rm GeV}$. Moreover,
\begin{equation}
Y=2(R_q-R_{q_{\rm ex}}+\tilde{R}_W),
\end{equation}
where $q_{\rm ex}$ denotes the exchanged quark in the t-channel. We have $R_q=m_q^2/m_\zeta^2\sim 1/4$ and $R_{q_{\rm ex}}=m_{q_{\rm ex}}^2/m_\zeta^2$. If we choose the exchanged quark to be a $b$ quark with the off shell mass, $m_{q_{\rm ex}}\sim 4.5\,{\rm GeV}$, we obtain the partial decay width:
\begin{equation}
\Gamma(\zeta\rightarrow WW^*)=1798\,{\rm keV}.
\end{equation}
The partial decay width for the Higgs boson is: $\Gamma(H\rightarrow WW^*)=875\,{\rm keV}$.
$$ $$
3. Decay channels $\tau^+\tau^-,b\bar b,c\bar c$
$$ $$
The quarkonium states decay into lighter fermion states via s-channel $\gamma, Z$ or via t-channel $W$ exchange. The initial states which decay into $f\bar f$ are restricted, because the diagrams are all s-channel exchanges. The photon contributes only to the $1^{--}$ quarkonium state, while the $Z$-exchange contributes to $1^{--}$ exchange through vector coupling and $0^{-+}, 1^{++}$ through axial vector coupling. The decays of $\zeta(0^{-+})$ and $\zeta(0^{++}),\zeta(1^{++}),\zeta(2^{++})$ to massless fermions are forbidden by chirality conservation. The partial widths of decays of the $\zeta(0^{-+})$ to $f\bar f$ are for the $^1S_0$ state~\cite{Barger}:
\begin{equation}
\Gamma(\zeta\rightarrow f\bar f)=\frac{12\pi\alpha_Z^2N_c\beta_f}{32}\frac{m_f^2}{m_Z^4}\vert R_S(0)\vert^2.
\end{equation}
Here, $N_c$ is the color factor equal to $1$ for leptons and $3$ for quarks, $\beta_f=(1-4R_f)^{1/2}$ denotes the velocity of the final fermion in the quarkonium rest frame with $R_f=m_f^2/m_\zeta^2$.

We obtain for the partial decay width of $\zeta\rightarrow\tau^+\tau^-$:
\begin{equation}
\Gamma(\zeta\rightarrow\tau^+\tau^-)=\frac{\pi\alpha_Z^2\alpha_s^3\beta_\tau
m_\tau^2m_\zeta^3}{18m_Z^4}.
\end{equation}
We use $m_\tau=1.777$ GeV and $\beta_\tau=(1-4R_\tau)^{1/2}=0.997$ and obtain
\begin{equation}
\Gamma(\zeta\rightarrow\tau^+\tau^-)=43.3\,{\rm eV}.
\end{equation}
We obtain the ratio:
\begin{equation}
{\cal R}_{\tau\tau}=\frac{\Gamma(\zeta\rightarrow\tau^+\tau^-)}{\Gamma(H\rightarrow\tau^+\tau^-)}=1.67\times 10^{-4}.
\end{equation}
This represents a significant suppression of the $\zeta$ resonance $\tau^+\tau^-$ decay rate compared to the Higgs boson decay rate. The present CMS signal strength for the $\tau^+\tau^-$ channel is consistent within 1 $\sigma$ with  the standard model or with $0$~\cite{CMS}.

The decay rate for $\zeta\rightarrow b\bar b$ is given by
\begin{equation}
\Gamma(\zeta\rightarrow b\bar b)=\frac{36\pi\alpha_Z^2\beta_bm_b^2\vert R_S(0)\vert^2}{32m_Z^4}=\frac{\pi\alpha_Z^2\alpha_s^3\beta_bm_b^2m_\zeta^3}{6m_Z^4}.
\end{equation}
We get the result
\begin{equation}
\Gamma(\zeta\rightarrow b\bar b)=0.72\,{\rm keV}.
\end{equation}
As with the $\tau^+\tau^-$ channel, we see that there is a significant suppression of the $\zeta$ resonance $b\bar b$ decay channel:
\begin{equation}
{\cal R}_{b\bar b}=\frac{\Gamma(\zeta\rightarrow b\bar b)}{\Gamma(H\rightarrow b\bar b)}=3.1\times 10^{-4}.
\end{equation}

Finally the decay rate $\Gamma(\zeta\rightarrow c\bar c)$ is
\begin{equation}
\Gamma(\zeta\rightarrow c\bar c)=68.3\,{\rm eV}.
\end{equation}

This predicted suppression of the $\tau^+\tau^-$, $b\bar b$ and $c\bar c$ decay channels compared to the Higgs predictions will constitute an important experimental way to distinguish the Higgs boson and the quarkonium $\zeta$ boson models.

We display in Table 1 the results of the calculations of the $\zeta$ and Higgs boson decay rates and branching ratios. The Higgs decay rates and branching ratios are from ref.~\cite{CERNworkinggroup}.

\begin{table}
\caption{$\zeta$ and Higgs boson decay rates}
\begin{center}
\begin{tabular}{|c|c|c|c|c|}
\hline
Decay channels & $\zeta$ decay rate& Higgs decay rate& $\zeta$ branching ratio& Higgs branching ratio \\
\hline
{$\gamma\gamma$} & 48.3\,{\rm keV} & 9.1\,{\rm keV} & $2.9\times 10^{-3}$& $2.2\times 10^{-3}$\\
\hline
{gg} & 14.2\,{\rm MeV} & 349\,{\rm keV}& 0.86& $8.6\times 10^{-2}$ \\
\hline
$Z\gamma$& 92.7\,{\rm keV}& 6.36\,{\rm keV}& $5.6\times 10^{-3}$& $1.6\times 10^{-3}$ \\
\hline
{$ZZ^*$} & 331\,{\rm keV} & 107\,{\rm keV}& $2.0\times 10^{-2}$& $2.6\times 10^{-2}$ \\
 \hline
{$WW^*$} & 1798\,{\rm keV} & 875\,{\rm keV}& 0.11& 0.21 \\
\hline
{$\tau^+\tau^-$} & 43.3\,{\rm eV} & 259\,{\rm keV}& $2.6\times 10^{-6}$& $6.4\times 10^{-2}$ \\
\hline
\rule{0pt}{2.3ex}{$b\bar b$} & 0.72\,{\rm keV} & 2.35\,{\rm MeV}& $4.4\times 10^{-5}$& 0.58 \\
\hline
{$c\bar c$} & 68.3\,{\rm eV} & 118\,{\rm keV}& $4.1\times 10^{-6}$& $2.9\times 10^{-2}$ \\
\hline
\end{tabular}
\end{center}
\end{table}

The $\zeta$ and Higgs bosons will be primarily produced at the LHC by gluon-gluon fusion. We expect that the gluon production of the quarkonium resonance boson will be suppressed compared to the Higgs boson production. The Higgs boson couples to two gluons through a loop, whereas the two gluons couple directly to the quarkonium bound state and this coupling involves potential non-perturbative QCD effects.  Moreover, the quarkonium state is a mixture of top and bottom quarks and the coupling of the gluons is more dominant for the $b\bar b$ state than the shorter lived $t\bar t$ state.   At leading order and in the narrow-width approximation, the production cross section for the $\zeta$ boson in $pp$ collisions is given by
\begin{equation}
\label{zetaprodcrosssection}
\sigma(pp\rightarrow \zeta+X)=\frac{\pi^2\tau_\zeta}{8m_\zeta^3}\Delta\sigma\Gamma(\zeta\rightarrow gg)\int^1_{\tau_\zeta} dx\frac{\tau_\zeta}{x}g(x,Q^2)g(\tau_\zeta/x,Q^2),
\end{equation}
where $g(x,Q^2)$ is the gluon parton distribution function, $\tau_\zeta=m_{\zeta}^2/s$, and $s$ is the pp collision energy squared.  The factor $\Delta\sigma$ is a correction factor for the gluon fusion $\zeta$ boson production cross section. The Higgs production cross section is to leading order~\cite{Spira}:
\begin{equation}
\label{Higgsprodcrosssection}
\sigma(pp\rightarrow H+X)=\frac{\pi^2\tau_H}{8m_H^3}\Gamma(H\rightarrow gg)\int^1_{\tau_H} dx\frac{\tau_H}{x}g(x,Q^2)g(\tau_H/x,Q^2),
\end{equation}
where $\tau_H=m_{H}^2/s$. The ratio of the $\zeta$ production cross section to that of the standard model Higgs boson production cross section for  $m_H=m_\zeta$ is to leading order:
\begin{equation}
\frac{\sigma(\zeta)}{\sigma(H)}=\frac{\Delta\sigma\Gamma(\zeta\rightarrow gg)}{\Gamma(H\rightarrow gg)}.
\end{equation}

We obtain from (\ref{zetaprodcrosssection}), (\ref{Higgsprodcrosssection}) and Table 1 the ratio of the $\zeta$ boson and Higgs boson production cross sections:
\begin{equation}
\frac{\sigma(\zeta)}{\sigma(H)}=40.68\Delta\sigma.
\end{equation}
The total widths obtained from Table 1. for the diphoton decays are $\Gamma_{\rm tot}(\zeta)=14.7$ MeV and $\Gamma_{\rm tot}(H)=4.07$ MeV, where for a spin 0 boson $P$, $\Gamma_{\rm tot}(P)=\sum_i\Gamma_i(P\rightarrow ab)$ is the total width. The branching ratios for the $\zeta$ and Higgs bosons are
\begin{equation}
{\rm Br}(\zeta\rightarrow\gamma\gamma)=2.9\times 10^{-3},\quad {\rm Br}(H\rightarrow\gamma\gamma)=2.2\times 10^{-3}.
\end{equation}
The ratio of the two branching ratios is ${\rm Br}(\zeta)/{\rm Br}(H)=1.32 $. The signal strength for the decay of a spin 0 boson P into two photons is given by
\begin{equation}
\sigma^{SS}_{\gamma\gamma}(P)=\sigma(P)Br(P\rightarrow\gamma\gamma).
\end{equation}
The ratio of the signal strengths for the $\zeta$ and Higgs boson decays into two photons is
\begin{equation}
\mu_{\gamma\gamma}\equiv\frac{\sigma(\zeta)}{\sigma(H)}
\frac{Br(\zeta\rightarrow\gamma\gamma)}{Br(H\rightarrow\gamma\gamma)},\quad \mu_{ZZ^*}\equiv\frac{\sigma(\zeta)}{\sigma(H)}\frac{Br(\zeta\rightarrow ZZ^*)}{Br(H\rightarrow ZZ^*)},\quad \mu_{WW^*}\equiv\frac{\sigma(\zeta)}{\sigma(H)}\frac{Br(\zeta\rightarrow WW^*)}{Br(H\rightarrow WW^*)},
\end{equation}
and
\begin{equation}
\mu_{Z\gamma}\equiv\frac{\sigma(\zeta)}{\sigma(H)}
\frac{Br(\zeta\rightarrow Z\gamma)}{Br(H\rightarrow Z\gamma)}.
\end{equation}

If we adopt the value $\Delta\sigma\lesssim 0.019$ we obtain
\begin{equation}
\label{photonexcess}
\mu_{\gamma\gamma}\lesssim 1.0,\quad\mu_{ZZ^*}\lesssim 0.6,\quad \mu_{WW^*}\lesssim 0.4,\quad \mu_{Z\gamma}\lesssim 2.70.
\end{equation}
The ATLAS and CMS experiments~\cite{CMS,ATLAS} have found that the signal strengths for the two photon decay channel is 1.5-2 times bigger than the value predicted by the standard Higgs boson model. The result (\ref{photonexcess}) is consistent with the experimental result within the expected errors in the computation of the branching ratios.

From Table 1, we can read off the results for the ratio of the decay widths of $WW^*$ and $ZZ^*$:
\begin{equation}
\frac{\Gamma(\zeta\rightarrow WW^*)}{\Gamma(\zeta\rightarrow ZZ^*)}=5.4.
\end{equation}
The corresponding ratio of decay widths for the 125 GeV standard model Higgs boson is
\begin{equation}
\frac{\Gamma(H\rightarrow WW^*)}{\Gamma(H\rightarrow ZZ^*)} = 8.2.
\end{equation}
Accurate experimental results for the decay channel widths and branching ratios, obtained at the LHC will provide a way to distinguish the $\zeta$ boson and Higgs boson models. The branching ratios for the $\zeta$ and Higgs boson decays into $Z\gamma$ cannot be distinguished at present using the current CMS and ATLAS results.

\section{Conclusions}

We have developed a model in which the newly discovered boson at the LHC with a mass $125-126$ GeV can be identified with a heavy quarkonium, spin 0 pseudoscalar resonance $\zeta^0$. By mixing the two states $\vert\zeta\rangle$ and $\vert\zeta'\rangle$ through a rotation angle $\phi\sim 36\,^{\circ}$, we obtain two heavy quarkonium states $\vert\zeta\rangle$ and $\vert\zeta'\rangle$ with the masses $m_{\zeta}\sim 125$ GeV and $m_{\zeta'}\sim 230$ GeV. The $\zeta^0$ can form a bound state with the standard QCD gluon interaction and the mixing of the $\zeta$
and $\zeta'$ mesons is strongly enhanced by a non-perturbative gluon interaction produced by an axial current $U_A(1)$ anomaly.

A critical way to distinguish between the composite quark-anti-quark $\zeta$ meson and a standard model scalar Higgs boson is a precise determination of the decay branching ratios and signal strengths of the new boson. The decay rates and branching ratios of the $\zeta$ boson have been calculated using the non-relativistic quarkonium potential model and QCD with a strong QCD coupling constant $\alpha_s=0.118$~\cite{pdg}. The fermion-antifermion decays of the $\zeta$ meson are suppressed compared to the Higgs boson decays. On the other hand, the decays of the $\zeta$ boson to $\gamma\gamma, Z\gamma$, $ZZ^*$ and $WW^*$ yield branching ratios and signal strenghs comparable to the branching ratios and signal strengths of the Higgs boson.  This guarantees that the $\zeta$ meson model can be consistent with the ``golden channel'' decay results $X\rightarrow\gamma\gamma, X\rightarrow ZZ^*\rightarrow 4\ell$ and $X-rightarrow WW^*\rightarrow \nu\nu\ell\ell$, which are the predominant decay channels for the claim that the LHC has discovered a new boson.

The decay rates and branching ratios for the decay channels $\zeta\rightarrow\tau^+\tau^-$, $\zeta\rightarrow b\bar b$ and  $\zeta\rightarrow c\bar c$ are significantly suppressed compared to the Higgs boson decay predictions, which is in accord with the current results obtained by the ATLAS collaboration~\cite{VMartin,Puigh}, while it is marginally consistent with the CMS data~\cite{Dutta,Puigh}. Unless further analyses of the 2012 data gives a resolution of the difference between the ATLAS and CMS collaborations, we will have to wait for the experimental results obtained at the LHC when the machine is running at 13 TeV with a larger integrated luminosity.

In the event that the new boson is not a Higgs particle but a bound state $\zeta$ resonance, then we must consider an alternative mechanism that breaks electroweak symmetry. Three possible models have been proposed~\cite{Moffat3,Moffat4,Moffat5}. Many alternative models have been published. In the local field theory model described in ref.~\cite{Moffat5}, there are no scalar boson modes and the masses of the $W$ and $Z$ bosons and fermions are produced by the quantum vacuum self-energies of the particles associated with a dynamical vacuum symmetry breaking mechanism.

\section*{Acknowledgements}

I thank Bob Holdom, Martin Green, Viktor Toth and Yanwen Shang for helpful discussions. This research was generously supported by the John Templeton Foundation and in part by the Perimeter Institute for Theoretical Physics. Research at the Perimeter Institute is supported by the Government of Canada through Industry Canada and by the Province of Ontario through the Ministry of Economic Development and Innovation.

\end{document}